\documentclass[conference]{IEEEtran}
\usepackage{alltt}
\usepackage{graphicx}
\usepackage{wrapfig}
\usepackage{lettrine} 
\usepackage{amsmath}  
\usepackage{amsthm}   
\usepackage{amssymb}  
\usepackage{textcomp} 
\usepackage{xcolor}   
\usepackage{url}      
\usepackage{enumitem} 
\usepackage{upgreek}  
\usepackage{program}  

\def\Hmax{$\overline{\vbox to 1.4ex{}\mbox{h}}$}

\newenvironment{code}{\begin{small}\begin{alltt}}{\end{alltt}\end{small}}
\newtheorem{theorem}{Theorem}

\makeatletter
\@ifpackageloaded{txfonts}\@tempswafalse\@tempswatrue
\if@tempswa
  \DeclareFontFamily{U}{txsymbols}{}
  \DeclareFontFamily{U}{txAMSb}{}
  \DeclareSymbolFont{txsymbols}{OMS}{txsy}{m}{n}
  \SetSymbolFont{txsymbols}{bold}{OMS}{txsy}{bx}{n}
  \DeclareFontSubstitution{OMS}{txsy}{m}{n}
  \DeclareSymbolFont{txAMSb}{U}{txsyb}{m}{n}
  \SetSymbolFont{txAMSb}{bold}{U}{txsyb}{bx}{n}
  \DeclareFontSubstitution{U}{txsyb}{m}{n}
  \DeclareMathSymbol{\aleph}{\mathord}{txsymbols}{64}
  \DeclareMathSymbol{\beth}{\mathord}{txAMSb}{105}
  \DeclareMathSymbol{\gimel}{\mathord}{txAMSb}{106}
  \DeclareMathSymbol{\daleth}{\mathord}{txAMSb}{107}
\fi
\makeatother

\def\QED{\raisebox{0.8ex}{\framebox{\kern0.2ex}}}

\title{Safe and Chaotic Compilation for\\
Hidden Deterministic Hardware Aliasing}
\author{\IEEEauthorblockN{Peter T. Breuer}
\IEEEauthorblockA{%
Hecusys LLC, GA, USA\\
Email: ptb@hecusys.com}
}

\begin{document}

\maketitle

\begin{abstract}
Hardware aliasing occurs when the same logical address can
access different physical memory locations.  This is a problem for
software on some embedded systems and more generally when hardware
becomes faulty in irretrievable locations, such as on a Mars Lander.  We
show how to work around the hardware problem with software logic,
compiling code so it works on any platform with hardware aliasing with
hidden determinism.  That is: (i) a copy of an address accesses the same
location, and (ii) repeating an address calculation exactly will repeat
the same access again.  Stuck bits can mean that even adding zero to an
address can make a difference in that environment so nothing but a
systematic approach has a chance of working.  The technique is extended
to generate aliasing as well as compensate for it, in so-called chaotic
compilation, and a sketch proof is included to show it may produce
object code that is secure against discovery of the programmer's
intention.  A prototype compiler implementing the technology covers all
of ANSI C except longjmp/setjmp.
\end{abstract}

\section{Introduction}
\label{s:Intro}

\noindent
{\em Hardware aliasing}   describes ``the
situation where, due to either a hardware design choice or a hardware
failure, one or more of the available address bits is not used in the
memory selection process.'' \cite{Barr98} The effect used to be familiar
to programmers and users alike, as the `DLL hell' that the
old 16-bit versions of Windows were prone to.  Dynamic linked libraries
(DLLs) were problematic for many reasons, but one was that different
versions of the same library loaded at the same memory address and all
applications referenced the in-memory copy.  So if one application
loaded one version of the library then another loaded another version,
the first application would unpredictably, as far as the program and
user were concerned, find itself in the second library's code.

DOS users truly experienced the situation in the raw, as the ubiquitous expanded
memory managers (for memory beyond 1MB) such as QuarterDeck's QEMM
\cite{Glosserman85} remapped memory so the video graphics (VGA) and
bootstrap (BIOS) code shared addresses with random access memory (RAM).
What a program accessed at runtime at a given address depended on the
memory manager heuristics.  In consequence, code had to use standard
sequences to trigger expanded memory reliably.  Nowadays, processors
such as the Raspberry PI-2 (all models) \cite{upton2014raspberry} still
dynamically share up to 1GB of RAM for general-purpose use
with the graphics processing unit (GPU).

Whatever the physical rationale, the symptom is the same in every case.
It is that what looks to programs to be the same memory address (the
{\em logical} address) sporadically accesses physically different
resources.  Since the memory management hardware in modern processors
maps all input/output (I/O) into one monolithic processor address space,
that may mean different regions of RAM or it may mean different
peripheral devices such as USB (`universal serial bus') and GPU.

Certainly, modern applications programmers are more familiar with {\em
software aliasing}, in which the same physical resource -- memory
location or peripheral device -- is accessible via different logical
addresses.  But hardware aliasing has not gone away so much as become
less relevant for the ordinary programmer as platforms
have evolved to present a normalised
view to software.  The paradigm of platform hardware that actually has
hardware aliasing through and through but presents `normally' to even a
systems programmer is {\em oblivious} RAM (ORAM)
\cite{ostrovsky1990,ostrovsky1996,ostrovsky2013}.  That is a secure RAM
solution (available in various forms since the 90s) in which internally
the memory contents are continuously randomly aliased and re-aliased to
frustrate cold boot attacks \cite{Gruhn2013} (freezing the RAM sticks to
make the electrical charge last longer for analysis; RAM chips require a
read-and-rewrite cycle to maintain their contents and ordinarily taking
them out of the processor would lose the contents immediately). But the
closer the programmer works to hardware 
the less complete may be the facade. On platforms which are both
resource-limited and relatively inaccessible, programmers have to work
around not only hardware that shares the same address space
but faults that can exacerbate the situation.  A stuck bit may
take one address line out of commission, forcing the hardware map
into a smaller shared address space.  Or a Mars Lander may suffer a
cosmic ray through the processor that causes the arithmetic logic
circuit to compute $1+1=3$.  One way to cope with that is to
rewrite programs to treat $2$ and $3$ as indistinguishable for the purposes of
arithmetic.  That means rewriting to work in arithmetic
modulo a suitable equivalence. That may be, for example here, a partition of
the number space into pairs $\{2,3\}$, $\{4,5\}$, etc., in
which $\{2m,2m+1\} + \{2n,2n+1\} = \{2(m+n),2(m+n)+1\}$ and $\{2m,2m+1\}
* \{2n,2n+1\} = \{2mn,2mn+1\}$. That equivalence should in general be
the least indiscriminatory that
makes the processor's faulty answers $x+y$ into a homomorphic image of
standard computer arithmetic.\footnote{
The simplest construction of a suitable equivalence is to set the
processor's wrong result $1+1=3$ equivalent to the correct result
$1+1=2$, with $2\equiv 3$, and then close under the implications
$x_1\equiv x_2 \land y_1\equiv y_2\to x_1+y_1\equiv x_2+y_2$, where the
$+$ is the correct one. Similarly for multiplication, etc. The
equivalence shown is $x\equiv y$ iff $\lfloor x/2\rfloor=\lfloor
y/2\rfloor$, with $2\equiv 3$.
} That entails losing one
arithmetic bit or more, since the only homomorphic images have sizes
that are factors of the original size $2^{n}$, which will be $2^m$ for
some $m < n$.

While $4$ and $5$ may denote the same value arithmetically, as addresses
they access different memory locations and a program rewritten to
accommodate a nontrivial arithmetic equivalence will experience hardware
aliasing.  So the cure for the computational mathematics engenders a
need to handle the consequential aliasing.  Since low-level programs in
particular use pointers heavily, there is no room for an ad hoc approach
to that.  Even high-level programming languages use pointers nearly
ubiquitously -- every object in Java is really a pointer, so copying an
object requires the clone() system call.  A simple copy will just be
another pointer to the same object.  Every variable in Fortran95 is a
pointer -- a subroutine call f(x) alters the value of its callers
variable x when the subroutine writes to x.  Arrays in C are passed as
pointers too.  Not only will the system-level programmers need a
reliable way to program around hardware aliasing because systems
programming is hard enough without new traps for the unwary, but
application-level programmers, the mathematicians and control engineers,
must be able to remain comfortably ignorant of the hazard.

It turns out that there is a
systematic way of generating code that always works in a hardware
aliasing environment
-- a compiler mechanism -- and this paper sets it out.  There has been
no other work on this topic since the present authors made an original
suggestion as to what to do in 2014 \cite{BB14c},
but experience has shown that suggestion to be too delicate for
practice (the idea was to set up different type classes for arrays -- in
one, array entries are uniquely accessed by walking a pointer in unit
increments up from the bottom of the array, in another, access is
uniquely by walking down from the top, in another, access is uniquely
via a single constant offset from the bottom of the array, and so on;
each class enforces a unique calculation for each array entry which
produces the same result each time, as discussed below, but the
programmer must choose what type to declare each array as and 
there is no proof that deeply
nested arrays of pointers to arrays of pointers, etc., will not give
rise to inconsistencies).  The technology reported in this paper is less
efficient computationally but it is robust and usable in practice by
programmers, and it has been deployed and tested -- c.f.  the HAVOC
compiler suite for {\sc ansi} C \cite{ansi99} 
at \url{http://sf.net/p/obfusc}.  The programmer is aware
only that they need to declare each pointer with a {\bf restrict} modifier.
That nominates a top-level (or local) memory zone that the pointer is to
range through.  Not declaring a zone means the compiler assumes the
pointer can point anywhere at runtime, which causes prohibitively large
object code to be generated -- gigabytes for a line of source
code -- so it behoves the programmer to declare a small zone.

The solution works because of an underlying determinism in processors.
Processors are mechanisms that are both designed to produce repeatable
results and also do so inherently because of their material. They
are quite high power electronic logic circuits, which makes it rare
for a faulty bit to be really random.  Even if the electrical value is
floating then that will give one determined value when tested
electrically in a determined way -- in a semiconductor circuit the test
is always either if the bit can supply current to ground or to the
positive rail and the answer for a floating line is `no' both ways.  The
logic circuit has been damaged so that it is different from the way it
was designed but it is still a logic circuit which responds to given
inputs with given outputs, in this hypothetical case producing a
floating bit in some conditions.  The functionality depends on the
circuit around the damaged area and it may be more complicated input to
output than always on or always off, but it should be deterministic.

The situation may be encapsulated as {\em
hidden deterministic} hardware aliasing, axiomatised
as follows with respect to the addresses produced by sequences of
processor operations:
\centerline{Axioms}
\begin{enumerate}[label={\arabic{*}.}]
\item
A machine code copy instruction copies the physical bit sequence
exactly, such that a copied address accesses the same memory location as
the original;

\item
repeating the same sequence of operations produces an address that
physically has exactly the same bit sequence and accesses the same location;

\item
logically different addresses always have physically different bit sequences.
\end{enumerate}
\medskip

\noindent
By `logically different' is meant different in terms of the intended
arithmetic.  In the example, $4$ and $5$ are not logically
different as they are equivalent in the equivalence relation with
respect to which arithmetic is well-defined, with $4\equiv 5$, but $5$
and $6$ are logically different and $5\not\equiv 6$ in the equivalence
relation.  The $4$ and $5$ can be thought of as different physical encodings
of the same logical number (here, $2$), while $5$ and $6$ are different
physical encodings of different logical numbers.

Axiom~1 (`faithful copy') implies a compiler can generate code that
copies an address for later use after writing through it, and the copy
can be trusted to retrieve the written value.  The address must not be
altered, not even by adding zero, as any arithmetic calculation at all
potentially alters the physical representation of the address as a
sequence of bits, which then fails to access the same memory location.
In the example given, $0$ is equivalent to $1$, so adding what
is arithmetically zero may in fact mean adding one in physical terms.

Axiom~2 (`repeatability') allows for some calculations on
addresses, so long as they are repeated exactly each time.  That is useful
because machine code instructions to read or write a memory location
generally take a base address $a$ in a register and adds a displacement
$d$ embedded in the instruction to get {\em effective address} $a{+}d$
for the access.  It is impossible to avoid the processor doing that one
addition, but it does not matter because the same calculation is
repeated at each access, with physically the same sequence of bits
resulting.

Axiom~3 (`no confusion') guarantees that the representations as physical
sequences of bits of what are logically different addresses do not step
on each other.  The addresses with bit sequences $5$ and $6$ in the
example represent logically different numbers (logically $2$ and
$3$ respectively).

The layout of this paper is as follows.  Section~\ref{s:HA} explains how
to compile to obtain code `safe against hardware-aliasing' in the most
general context.  The generated code has to cope with variations in the
addresses produced, while they are actually governed by a `hidden
determinism', as described above.  Section~\ref{s:EC} details an
extension in which variations are deliberately introduced by the
compiler (which can cope them) to the maximal extent possible.  That has
significance for `obfuscation' in the object code produced.  The section
finishes with an argument that unrolling code to a length that is
exponential (super-polynomial) in the number of bits per word and then
compiling it with variations `to the maximal extent possible' as
described in the section results in safety against polynomial complexity
attacks aimed at determining what the data in the runtime trace means.
The technique introduced in Section~\ref{s:HA} can be seen as coping
with apparently unreliable {\em addressing} (that is really
deterministically generated but in an unknown way) and the technique of
Section~\ref{s:EC} extends that to generate apparent unreliability in
{\em data} in general, as well as cope with it.

\section{Safe Compilation for Hardware Aliasing}
\label{s:HA}

\noindent
The working principle for generating viable code in this kind of
environment is that each address that is written should be saved for
later read, as per Axiom~1 (`faithful copy') of Section~\ref{s:Intro}.
The problem is that it is saved at an address, which must also be saved,
and so on recursively.  Axiom~2 (`repeatability') allows a backstop to
be put on the recursion, via base addresses that are produced at compile
time and a repeated calculation at runtime based on them.  But no finite
set of addresses can suffice for nested function calls to unbounded
depth, so the runtime stack must be involved.  The first problem
is how to manipulate the {\em stack pointer} so addresses and
other data might be saved and recovered reliably from stack.

\subsection{Stack Pointer 101}

\noindent
The standard compiler-generated function call sequence
decrements the stack pointer
register {\bf sp} by the amount that will be
needed for local storage in the function immediately on entry to the
function body, and increments it again before exit:
\begin{equation}
\begin{minipage}{0.5\columnwidth}
\begin{code}
\rm call to{\em function}
 \dots
{\em function} code start:
{\color{red}\rm decrement}{\bf sp}
 \dots
{\color{red}\rm increment}{\bf sp}
\rm return
\end{code}
\end{minipage}
\label{e:1}
\end{equation}
That does not work in a hardware aliasing environment, because the
increment does not necessarily restore exactly the same physical
representation originally in the stack pointer register. Instead,
the caller gets back a possibly different set of bits that, however,
means the same thing arithmetically.  Being different, it references a
different location in memory.

The {\em frame pointer} register {\bf fp} must be co-opted to
copy the stack pointer, and copy the stack pointer
back from it just before return from the call:
\begin{equation}
\begin{minipage}{0.5\columnwidth}
\begin{code}
\rm{\it function} code start:
{\color{red}\rm copy}{\bf sp} to{\bf fp}
\rm decrement{\bf sp}
 \dots
{\color{red}\rm copy}{\bf fp} to{\bf sp}
\rm return
\end{code}
\end{minipage}
\label{e:2}
\end{equation}
That is the standard unoptimised function call sequence from a compiler
but optimisation will replace it with the {\bf fp}-less
code \eqref{e:1}.  The GNU {\em gcc} compiler (for
example) with {\bf-fno-omit-frame-pointer} on the command line turns off
optimisation and produces the code \eqref{e:2},
which works with hardware aliasing.

It is not perfect, however, because the caller's frame pointer
register must also be saved and restored by the callee around its
own use of the frame pointer register, as follows:
\begin{equation}
\begin{minipage}{0.5\columnwidth}
\begin{code}
\rm{\it function} code start:
{\color{red}\rm save} old{\bf fp} to 1 below{\bf sp}
\rm copy{\bf sp} to{\bf fp}
\rm decrement{\bf sp}
 \dots
\rm copy{\bf fp} to{\bf sp}
{\color{red}\rm restore} old{\bf fp} from 1 below{\bf sp}
\rm return
\end{code}
\end{minipage}
\label{e:3}
\end{equation}
Saving below the caller's stack pointer would ordinarily intrude on the
callee's stack area (`frame'), so the decrement must be larger by one in order
to leave room for it.  As many as the compiler
wants of the caller's registers can be saved 
like this.  The application binary interface (ABI) document for
the platform specifies which registers the callee must save, and which
the caller code must expect may be trampled on and must save itself.
The frame pointer and stack pointer are callee-saved in \eqref{e:3}.

The function call code \eqref{e:3} works well with hardware aliasing.
It allows local variables for the function to be reliably addressed as
${\it sp}{+}d$ on the stack, where $d$ is a displacement between 0 and
the function frame size.  The $d$ is supplied as an embedded constant in
a {\em load} or {\em store} machine code instruction (see below in
(\ref{e:4}-\ref{e:5})) that references the stack pointer register,
containing {\em sp}, as the base for the displacement.  The processor
calculates ${\it sp}{+}d$ for the {\em effective address} passed to
memory.

\subsection{Variables}
\label{ss:Var}

\noindent
With the function call sequence in \eqref{e:3}, accessing
local variables is simple.  A word-sized local variable {\bf x}
is assigned a position $n$ on the stack and the compiler issues a
load instruction ({\bf lw}) to read from there to register $r$:
\begin{equation}
\begin{minipage}{0.5\columnwidth}
\begin{minipage}{1.5\textwidth}
\begin{code}
{\color{red}lw}{\it r}{\it n}({\bf{sp}})  \#\rm load from offset \(n\) from{\em sp}
\end{code}
\end{minipage}\hss
\end{minipage}
\label{e:4}
\end{equation}
The processor calculates ${\it sp}{+}d$ 
but repeats the same calculation at every access, so by Axiom~2
of Section~\ref{s:Intro} the same sequence of bits for 
the address is produced every time, and it accesses the same spot in
memory.  To write the variable, a {\em store} instruction
({\bf sw}) replaces the load instruction:
\begin{equation}
\begin{minipage}{0.5\columnwidth}
\begin{minipage}{1.5\textwidth}
\begin{code}
{\color{red}sw}{\it n}({\bf{sp}}){\it r}  \#\rm store to offset \(n\) from{\em sp}
\end{code}
\end{minipage}\hss
\end{minipage}
\label{e:5}
\end{equation}
For a global variable  at an address $a$
in (heap) memory, the compiler offsets from the zero register {\bf zer}
instead:
\begin{equation}
\begin{minipage}{0.5\columnwidth}
\begin{minipage}{1.5\textwidth}
\begin{code}
lw{\it r}{\it a}({\color{red}\bf{zer}})  \#\rm load from address \(a\)
\end{code}
\end{minipage}\hss
\end{minipage}
\end{equation}
The zero register contains a fixed base value.  The  effective
address sent to memory is $a{+}0$, which is possibly a
physically different sequence of bits to $a$ (representing the same value),
but the calculation is repeated exactly each time
so the same memory location is accessed each time.

Variables in the parent's frame may also be accessed.  If the function
is defined within another function, we shall call it an {\em interior}
function (`nested' is the standard term, but it risks confusion with
`nested' function calls -- one function called from another).
The compiler arranges that the exterior function's frame pointer is
handed down at runtime in the {\bf c9} register (it is {\bf c9}
in our own platform's
ABI; it may be different in other ABIs) and  that is preserved
through successive interior function calls.  Then a load or store
instruction using {\bf c9} instead of {\bf sp} or {\bf zer} reliably
accesses the exterior function's variables on the stack.

\subsection{Arrays}

\noindent
Arrays present the real difficulties as entries are fundamentally always
accessed by address and addressing is fundamentally unreliable in a
hardware aliasing environment.

There are (at least) two common but different ways of addressing the
entries in an array $\bf a$.  That is (a) via a load or store
instruction with fixed displacement $n$ from the array address $a$ (that
is `$a[n]$'), or (b) via a pointer with value $p$ that ranges through
the array starting at $a$ and steps through the elements until the
desired one is reached, at which point a load or store instruction with
displacement 0 from the pointer (`$p[0]$') is applied.  Those two modes
were considered in our earlier work \cite{BB14c},
which continues to
be the only work we know of on how to even possibly get around hardware
aliasing.  The two calculations (a) and (b) for the effective address
are respectively $a+n$ and $a+1+1\dots+1+0$.  The calculations may
produce physically different sequences of bits to be sent to memory, so
the two are mutually incompatible and one or the other must be used
all the time for consistency.  But both have proven too restrictive in
practice.  It is as common, for example, for real code to step a pointer
down through an array as to step up through it, and the transformation
the compiler needs to do results in prohibitively inefficient code at
runtime. Evidently also, there are many other ways of addressing array
entries commonly encountered in source code and we cannot be expected
to cater for  them all.

We have accepted instead an engineering compromise in which array access
is not for general-purpose use going to be constant time.
For special purposes, one of the addressing modes (a)
and (b) may be used, but that would be liable to cause programming mistakes
at application-level.  For arrays of
size $N$ it turns out the compiler can provide access in $\log N$ time in a
simple manner that is safe and reliable for all 
methods of calculating an index or pointer, known or unknown.

Linear complexity code will be presented first.
To read (local) array element $\bf a[n]$ the code
tests $\bf n$ against each of $0,\dots,N{-}1$ in turn and chooses
one address for each entry:
\begin{equation}
\begin{minipage}{0.5\columnwidth}
\begin{code}
(n == 0)?a[0]:
(n == 1)?a[1]:
\dots
\end{code}
\end{minipage}
\end{equation}
The equality tests are arithmetic and are therefore insensitive to the
physical representation of the value $n$ of {\bf n} as possibly physically
different sequences of bits.  The generated machine code always passes
effective address $a+d$
to memory where $d$ is the displacement from the base of the array for the
entry and $a$ is the address ${\it sp}+k$, where $k$ is the position on the
stack allocated by the compiler for the lowest array element $\bf a[0]$. The
code is just the one machine instruction:
\begin{equation}
\begin{minipage}{0.5\columnwidth}
\begin{minipage}{1.5\textwidth}
\begin{code}
lw{\it r}{\it d}({\it{r}})     \#\rm load from address \(a{+}d\) 
                     \# with \(a\) in{\it r} to{\it r}
\end{code}
\end{minipage}\hss
\end{minipage}
\label{e:8}
\end{equation}
The address $a$ is supplied by a preceding instruction:
\begin{equation}
\begin{minipage}{0.5\columnwidth}
\begin{minipage}{1.5\textwidth}
\begin{code}
{\color{red}addi}{\it r}{\bf sp} \(k\)   #{\rm add \(k\) to{\it sp} in{\it r}}
lw{\it r}{\it d}({\it{r}})     #\rm load from address \(a{+}d\)
\end{code}
\end{minipage}\hss
\end{minipage}
\end{equation}
That produces the same calculation ${\it sp}{+}k$ for $a$ every time.

Improving this code to $\log N$ complexity means using a binary tree
instead of linear lookup for the value $n$ of {\bf n},
deciding first if $n$ is below $N/2$ or above it, then on what side of
$N/4$ or $3N/4$ it is, and so on.
Code for writing instead of reading follows the same pattern, with store
instead of load instructions at the leaves of the binary tree or linear
sequence.

The same form works for pointer {\bf p} accesses, provided
the compiler knows what zone of memory it points into.  We have
tightened the type system of C (the source language for our prototype
compiler) so the pointer is declared with the name of a (possibly over-large)
array {\bf a} into which the programmer guarantees it points at runtime:
\begin{equation}
\begin{minipage}{0.5\columnwidth}
\begin{code}
int *p{\bf restrict} a
\end{code}
\end{minipage}
\end{equation}
That selects array {\bf a} as the target zone for {\bf p}.

Some porting has to be done for existing code, marking out areas
into which different pointers point.  It generally turns out to mean
declaring a global array from which objects of the kind pointed to are
allocated from, or declaring one function as interior to another
function (for example, main) where the target zone of the pointer is
declared as local on the stack.

A {\bf restrict} pointer type is narrower than it would otherwise be but
there are no no semantic changes to the language so the programmer does
not have to relearn anything.  The programmer does (usually) have to
make changes, but the new code is valid in unmodified C, so it can be
checked.  Ideally it is obtained via a sequence of careful code
transformations.  The conservative choice is usually to replace ad-hoc
declarations with calls to an object factory for each compound type.

The following code is suitable for lookup via pointer {\bf p}:
\begin{equation}
\begin{minipage}{0.5\columnwidth}
\begin{code}
(p == a+0)?a[0]:
(p == a+1)?a[1]:
\dots
\end{code}
\end{minipage}
\label{e:11}
\end{equation}
It is insensitive to the way the pointer $\bf p$ is calculated because
the equality tests are arithmetic and are not derailed by the aliasing
in memory addressing. 
A selection $\bf a[d]$ is made, and the same lookup code \eqref{e:8} for $\bf
a[d]$ is executed each time, giving the same result.

This code can (also) be made $\log N$ complexity with a binary
tree structure.  It can be converted for write by replacing load
instructions at the leaves with store instructions.

These constructions make pointer-based addressing consistent with access
via an array index in the hardware aliasing context. The idea is to 
make the choice of address arithmetically, and then reuse that same
address or calculation for the address again and again.

If the number {\small\bf N} of array elements is determined 
at runtime (the C '99 standard adds so-called
{\em variable length arrays}),
then a `late binding' code pattern is needed instead.
The following generated code generalises 
\eqref{e:11} to read reliably through
pointer {\bf p} for a number {\small\bf N} of elements that is
a local variable at runtime:
\begin{equation}
\begin{minipage}{0.5\columnwidth}
\begin{code}
for (int d=0; d<{\bf{N}}; d++)
  if (p == a+d)\,\{\,return a[d];\,\}
\end{code}
\end{minipage}
\label{e:12}
\end{equation}
The effective address passed to memory is $sp+k+(0+1+\dots+1)+0$, where
$a={\it sp}+k$ is the address of array {\bf a}, and the 1s are repeated
$d$ times to address the $d$th element of the array. This is
not the same calculation as in \eqref{e:11} so it is not generally
compatible with that in this kind of environment. The compiler must
always generate either the
form \eqref{e:11} or the form \eqref{e:12}. If there are no dynamically
sized arrays in the code then it can afford to use \eqref{e:11} (the
generated code is longer but conceptually simpler).  If there are
dynamically sized arrays then it must use the form \eqref{e:12}.
Tighter distinctions may be possible, but we have not explored them.

The same form must be used for indexed read:
\begin{equation}
\begin{minipage}{0.5\columnwidth}
\begin{code}
for (int d=0; d<N; d++)
    if (n == d)\,\{\,return a[d];\,\}
\end{code}
\end{minipage}
\label{e:13}
\end{equation}
The calculation for the effective address passed to memory is the same
as in \eqref{e:12}.
Both (\ref{e:12}-\ref{e:13}) can be modified for $\log$\,N complexity,
and then the modified form must be used always.

Writes replace the `$a[d]$' return (translated to a load machine code
instruction by the compiler) with a `$a[d]=x$' (translated to a store
machine code instruction).

\subsection{Data Types}
\label{ss:MW}

\noindent
We will not unduly belabour the topic here, but compound (and short)
data types need special consideration, at least because compound data
may contain arrays and arrays need special consideration (above).
Indeed, a common programming style in C is to access the members of a
compound data type as though the whole were an array of words, and to
access the members via a pointer to word.  Unless debugging memory
pointer problems is a favourite pastime, the programmer wants the
compiler to get both modes of access (conventional and as an array of
words) right.

The trick for the compiler is to always treat records with named
fields (`struct' in C) as word arrays and translate the
field name to an array displacement.  The declaration
\begin{equation}
\begin{minipage}{0.5\columnwidth}
\begin{code}
struct \{ int a; int b; \} x
\end{code}
\end{minipage}
\end{equation}
declares {\bf x} with
two named fields, {\bf a} and {\bf b}, each one word wide. It occupies
two words on the stack at displacements $k$ and $k'$ (the value
$k{+}1$)
respectively from the stack pointer. The compiler generates accesses to
the fields
{\bf x.a} and {\bf x.b} just as it would for any local variables
situated there, by calling
\begin{equation}
\begin{minipage}{0.5\columnwidth}
\begin{code}
lw{\it r}{\it k}({\bf{sp}})  \#\rm load from {\bf x.a} 
\end{code}
\end{minipage}
\end{equation}
to read from {\bf x.a}, for example. The effective address passed to memory
is ${\it sp}{+}k$. To access {\bf x.b}, the address passed is ${\it
sp}{+}k'$ instead.  The compiler generates the code for array access
explained in the section above, and source code that accesses the fields
of the struct as though it were an array works too.

Long atomic types such as {\bf double} are also treated as arrays of words
by the compiler.  But most platforms have double-word
load and store instructions that will fetch/write two words at once:
\begin{equation}
\begin{minipage}{0.5\columnwidth}
\begin{code}
ld{\it r}{\it k}({\bf{sp}})  \#\rm double word load 
\end{code}
\end{minipage}
\end{equation}
and only the address of the first word is given to the instruction.  Registers
are indexed as pairs for this instruction, and the partner to $r$ is loaded
up by the instruction with the word at address $k{+}1$.

But that is not necessarily compatible with treating the double as a
two-word array in a hardware aliasing environment.  If the double is on
the stack then the effective address for the first word is ${\it
sp}{+}k$, say.  The processor may request a double-word from memory at
that address.  Or it may request two words, one at ${\it sp}{+}k$ and
one at $({\it sp}{+}k){+}1$.  Both address calculations are reliably
repeatable, according to the axioms of Section~\ref{s:Intro}, but by
those axioms there is no guarantee that the latter calculation accesses
the word at one beyond ${\it sp}{+}k$ in memory.  Indeed, the example
given in that section shows that if ${\it sp}{+}k$ gives one of the
answers $\{4,5\}$, then the arithmetic must be so arranged that $({\it
sp}{+}k){+}1$ gives one of the answers $\{6,7\}$.  It may be that $4$ is
produced for the former, and $6$ for the latter, and those are not
consecutive positions in memory.

Overall, it is safer that the compiler {\em not} make use of double word
instructions, though if the program contains no accesses to doubles as
two-word arrays, then it is perfectly safe.

The difficulty remarked above with {\bf double} transfers to accessing
the individual bytes of a single word too.  It is safer overall not to
use the byte-oriented instructions that the platform provides, but to
use arithmetic instead, as follows.

For index-oriented access to the characters of a string {\bf a},
the compiler generates code that splits the character index {\bf i}
into index $\bf d$ for a word consisting of a sequence of 4 characters,
and offset $\bf j$ for the wanted character within the word:
\begin{equation}
\begin{minipage}{0.5\columnwidth}
\begin{code}
d = i/4; 
j = i%4;
\end{code}
\end{minipage}
\end{equation}
Then the character  is obtained via an array-of-words lookup and 
the following arithmetic for the \(j\)th char of the $d$th word:
\begin{equation}
\begin{minipage}{0.5\columnwidth}
\begin{code}
(a[d] / 256\(\sp{\tt{j}}\)) % 256 
\end{code}
\end{minipage}
\end{equation}
In our own prototype compiler, we have
preferred to avoid the complication and pack characters only one to a
word, at the cost of an inefficient use of memory.

\section{Deliberately Chaotic Compilation}
\label{s:EC}

\noindent
We can take the general scheme for compiling around hardware aliasing on the
platform and use it to work around variations introduced by the compiler
itself.  The point is that varying the object code tends to obscure what
it does, and obfuscation is a legitimate aim of some fields of software
engineering. Many companies might like to distribute object code that is
so obfuscated that it cannot be reverse engineered, while not being
hampered in execution (of course it is easy to produce code that
theoretically cannot be reverse engineered -- it suffices to include a
loop that in parallel to everything else that the program does, searches
at low priority for an inconsistency proof for integer arithmetic, and
if it finds one, interrupts the rest of the program, and nobody can say
for certain if the program will complete normally or not -- but in
practice a skilled engineer can make a good job of it).

The aim in chaotic compilation is to compile the same source code 
differently each time, forming object codes that have the same overall
functionality but the runtime traces vary to the maximal extent possible
within these constraints:

\begin{enumerate}
\item
The same sequence of machine code instructions executes, in the same
order.
\item
The constants embedded in each machine code instruction differ
to the maximal extent possible.
\item
The data written to registers and memory differs to the maximal extent possible.
\end{enumerate}

\noindent
The idea is to compile program $p_1$ and prepare data $d_1$ for it,
and to compile program $p_2$ and prepare data $d_2$ for it. The
two data sets represent the same thing, but their presentation for the
two compiled programs differs.  Then $p_1(d_1)$ and $p_2(d_2)$ are run.
Comparing the two runtime traces, the characteristics 1)-3) above are
seen.  The `differs to the maximal extent possible' refers to the
stochastic distribution of outcomes.  Compilation is randomised, so each
runtime trace is one of a range of possible traces, and the ideal is
that each trace is exactly as frequent an outcome as any other among the
possibilities.  That is a `flat' distribution of outcomes, which is a
maximal entropy\footnote{The entropy of a distribution of random
events $X$ is formally the expectation $E[-\log_2 {\rm prob}(X)]$.
Informally, it captures the number of 1/0 degrees of freedom (`bits').
Entropy of 1 bit equates to two equally probable possibilities $X$.}
distribution in information theoretic terms.
That is, no bias or other tendency is discernible.  Even though human
programmers naturally tend to use low numbers like 0 and 1, the compiler
will have randomised the object code so that tendency, and any other, is
not present.  So `differs to the maximal extent possible' means a flat
probability distribution of the stated observations, which is
technically {\em chaotic}.

`Getting the intended result' would be a fourth constraint, but it is
already said that the overall functionality is retained.

\def\Ob{$\cal O$}

It will be argued that
chaotic compilation may be applied to any program
such that on a platform with a $n$-bit word:
\begin{equation}
\begin{minipage}{0.85\columnwidth}
The runtime trace cannot be read correctly with 
probability above chance by a polynomial time method
\end{minipage}
\tag{\Ob}
\label{obfusc}
\end{equation}
in $n$ as $n\to\infty$, in the (hypothetical) situation that the
hardware word size can be varied.  That is, the probability $p$ of
getting right what it is intended by the programmer for any chosen data bit
in the trace to mean tends to $1/2$ as $n\to\infty$.

In the simple
case of a program with no code, for example, the compiler produces no
machine code for both programs $p_1$ and $p_2$ above.  The prepared data
$d_1$ and $d_2$ still differ, however.  For example, in $d_1$ the author
may write $3$ where they really mean $7$, but in $d_2$ they may write
$15$ where they really mean $7$, and so on.  The author can still read
the output (which is the same as the input) because they know the
substitution.  But an onlooker cannot tell what they really meant.
There is no bias in the observations if the presentation scheme
for the input data (which is the same as the output) is randomly chosen
(this is a fundamental result of information theory -- the entropy in
two $n$-bit signals added together cannot be less than the entropy of
either, so when one has maximal entropy, the combined signal also has
maximal entropy, which means a flat distribution).

The more remarkable part is that the reasoning works for any program.
We will show how the compiler may vary
\begin{enumerate}[label={{\it\Alph*}.}]
\item addresses \quad(Subsection {\em\ref{ss:ADC}} below);
\item data content (Subsection {\em\ref{ss:CD}})
\end{enumerate}
randomly so traces `differ to the maximal extent possible' at runtime
between compilation and recompilation, elaborating the technique 
of Section~\ref{s:HA} to induce variation as well as cope with it,
concluding with a sketch of the argument for \eqref{obfusc}.

\subsection{Address Displacement Constants}
\label{ss:ADC}

\noindent
Instead of generating a load instruction to read from a variable at
position $n$ on the stack like this \eqref{e:8}:
\begin{center}
\begin{minipage}{0.5\columnwidth}
\begin{minipage}{1.5\textwidth}
\begin{code}
lw{\it r}{\it n}({\bf{sp}})  \#\rm load \(r\) from offset \(n\) from{\em sp}
\end{code}
\end{minipage}
\end{minipage}
\end{center}
the compiler will issue the instruction with a {\em
displacement constant} $\Delta$ different from $n$:
\begin{equation}
\begin{minipage}{0.5\columnwidth}
\begin{minipage}{1.5\textwidth}
\begin{code}
lw{\it r} {\color{red}\(\Delta\)}({\it{s}})  \#\rm load from offset \(n\) from{\em sp}
\end{code}
\end{minipage}
\end{minipage}
\label{e:19}
\end{equation}
$\Delta$ is randomly chosen and the register $s$
has been pre-set to contain ${\it sp}+n-\Delta$,
where {\it sp} is the nominal value of the stack pointer. The bit sequence 
passed to memory by \eqref{e:19} as effective address to read from
is the result of the calculation:
\begin{equation}
{\it sp} + n - \Delta + \Delta
\label{e:20}
\end{equation}

\noindent
The compiler always emits the same instruction sequence, getting the same
address always as result, but it has to ensure that the $\Delta$ used
in place of the $n$ is the same each time, for each $n$. So
it maintains a vector ${\bf\Delta}$ indexed by stack location $n$.
(a similar vector ${\bf\Delta}_Z$ is maintained for the heap). Then
the $\Delta$ in \eqref{e:19} is really ${\Delta=\bf\Delta}n$ and the
instruction in \eqref{e:19} is:
\begin{equation}
\begin{minipage}{0.5\columnwidth}
\begin{minipage}{1.5\textwidth}
\begin{code}
lw{\it r} {\color{red}\({\bf\Delta}{n}\)}({\it{s}}) #{\rm load from offset \(n\) from{\it sp}}
\end{code}
\end{minipage}
\end{minipage}
\label{e:21}
\end{equation}
The vector ${\bf\Delta}$ is changed randomly by the compiler as it works
through the source code and is a source of extra variation, but as
explained above it seems not to do anything because the base address
read from register $s$ is asserted in \eqref{e:20} to be offset to
compensate.  That is not true.  The compiler controls the offset in 
register $s$ too, as explained below, so that random
variations from nominal in every register and memory location are
produced and accounted for. The essence is that:
\begin{equation}
\begin{minipage}{0.8\columnwidth}
\em Each machine code instruction that writes 
is freely varied to the maximum extent possible.
\end{minipage}
\tag{\Hmax}
\label{hmax}
\end{equation}
For example, the load instruction of \eqref{e:21} has one embedded constant
${\bf\Delta}n$ and it is freely varied by the compiler. The instruction is 
exceptional, however, in that it writes, but not in any way that is
controlled by the varied constant.  The result always ends up in the
named register $r$.  Instead `writes'
must be understood as `having an observable effect on the trace'.

\subsection{Content Deltas}
\label{ss:CD}

\noindent
As remarked above, the stack pointer {\bf sp} does not contain the
value {\it sp} that it notionally should have but instead
is offset from that by a randomly generated value $\delta$.
That is true of the content of every register and memory location
at every point in the generated
code.  The compiler maintains a vector ${\bf\eth}$ of the offset
delta for content in each register and memory location, varying it as it
passes through the code, and the $\delta$ for the stack pointer register
is $\delta = {\bf\eth}\,{\bf sp}$.

We need to work through enough detail of what the compiler does to show
the principle \eqref{hmax} is satisfied, but most will be omitted.  The
following is an abstract, declarative rendering of what the compiler
does to translate a non-side-effecting expression $e$ of the source
language, which is the simplest part of its work.

Let the compiler be $C_r[-]$, translating $e$ to machine code {\it mc} 
that targets the result for register $r$ at runtime. That is:
\begin{align}
({\it mc},{\bf\eth}) &= C_r[e] 
\label{e:22}
\end{align}
Let the state of the runtime machine before {\it mc}
runs be $\sigma_0$, let the nominal value for the expression\footnote{
The `nominal value' $[e]^{\sigma}$ of expression $e$ is 
is formalisable via a canonical construction: map a
variable $x$ in the expression to its register location $r_x$ (the
runtime value is offset by a delta ${\bf\eth} {r_x}$), checking the
content of $r_x$ in the state and discounting the delta to get
$[x]^{\sigma}=\sigma(r_x)-{\bf\eth} {r_x}$.  Arithmetic 
in the expression is formalised recursively, with
$[e_1+e_2]^{\sigma}=[e_1]^{\sigma}+[e_2]^{\sigma}$, etc.
} be $[e]^{\sigma_0}$, then
running code {\it mc} takes state $\sigma_0$ to state $\sigma_1$ in which the
value in register $r$ is offset from the nominal value by the 
randomly generated amount ${\bf\eth} r$. That is:
\begin{align}
\sigma_0 &\stackrel{\it mc}{\rightsquigarrow} \sigma_1
~~{\rm where}~ \sigma_1(r) = [e]^{\sigma_0} + {\bf\eth} r
\label{e:23}
\end{align}

\noindent
Re-rendering the code in \eqref{e:21} to read 
the $n$th location on the stack and show explicitly
the previous instruction
that preps the base address (in register $r$) for the load:
\begin{equation}
\begin{minipage}{0.5\columnwidth}
\begin{minipage}{1.5\textwidth}
\begin{code}
addi r{\bf sp} \(k\)  \#{\rm \(k=n-{\bf\eth}\,{\bf{sp}}-{\bf\Delta}{n}\)}
lw{\it r} \({\bf\Delta}{n}\)({\it{r}})  \,\,\#{\rm read \(n\)th stack location}
\end{code}
\end{minipage}
\end{minipage}
\label{e:24}
\end{equation}
The memory receives as effective address to read from 
the result of the following calculation:
\begin{equation}
{\it sp} + {\bf\eth}\,{\bf sp}  + k + {\bf\Delta}n
~\text{where}~ k = 
n-{\bf\eth}\,{\bf{sp}}-{\bf\Delta}{n}
\label{e:25}
\end{equation}
The stack pointer register {\bf sp} contains the nominal value ${\it sp}$
offset by ${\bf\eth}\,{\bf sp}$.  Summing, the address has the
arithmetic value ${\it sp}{+}n$. The physical representation as a
sequence of bits may both be different from nominal and
may vary, as discussed in Section~\ref{s:Intro}: one might equally get a
$4$ as a $5$ from the calculation, following the example there.
However, the calculation is the same every time so the bit sequence
passed to memory is the same every time, by Axiom~2.  It cannot hit the
memory location associated with any other (logical) address by Axiom~3,
and the calculation for the address \eqref{e:25} is repeated
exactly at every access via \eqref{e:24} so it hits just one
memory location of those feasible for the logical address.

The $\bf\eth$ and $\bf\Delta$ values are changed by the compiler at
(just before) every point in the code where a write occurs.
After the write the deltas for that location have to be maintained
constant through the following sequences of reads from the same location
along the code paths through that point, or the reads would miss.  There
are two instructions generated in \eqref{e:24} and both carry constants
that can be varied by the compiler.  However, the constraint on the
right in \eqref{e:24} binds them and restricts the total variation
possible.  A further variation beyond that constraint would to 
vary the logical position $n$ of the target on the stack at every write
to it.  That can be done but it is too tricky to describe here, so we
will hypothesise that the position on the stack forms part of
the intended program semantics and it cannot be varied.

Within that constraint, the two instructions in \eqref{e:24} can
be varied maximally by the compiler, as required by \eqref{hmax}.
It is just that, by virtue of the constraint, the entropy the compiler
can introduce to the runtime trace via them is not $2\times$ 32
bits (assuming a 32-bit word), but only $1\times$ 32 bits. That is, if the
${\bf\Delta}n$ constant in the load instruction is freely chosen by the
compiler, then the $k$ constant in the preceding addition is determined
by $k = n-{\bf\eth}\,{\bf{sp}}-{\bf\Delta}{n}$, and if the $k$ constant
is freely chosen then the ${\bf\Delta}n$ value is determined.

Closely related
work presently under review \cite{cryptoeprint:2019:084}
makes the above observations precise:
\begin{theorem}
The entropy in a trace over recompilations is the sum of the entropies
of every instruction that writes that appears in it, counted once
each.
\label{t:1}
\end{theorem}

\noindent
The entropy of the two instructions combined in \eqref{e:24} is 32 bits
(on a 32-bit platform).  That can be seen as 32 bits for the first
instruction, and zero for the second, as its variation is already
determined once the first has been seen.

The compiler's job in the chaotic compilation context
is to do everything it can to maximise entropy in the trace. It turns
out that following \eqref{hmax} as a compiler design principle does
that, which is why we have been careful to check for it above:
\begin{theorem}
The trace entropy is maximised when the compiler varies every
instruction that writes to the maximal extent possible
from recompilation to recompilation.
\label{t:2}
\end{theorem}

\noindent
Successfully varying each instruction to the maximal extent possible
provides a stochastic setting at runtime in which an observer cannot be
sure what the numerical value of the data in the trace is really
intended to mean even in terms of a statistical tendency, because the
maximum 32 bits (on a 32-bit platform) of information caused to be
written by the programmer at each point in the trace is swamped by 
that much contribution again from the compiler.  But there are
limits on what the compiler can do in terms of variation,
as argued for \eqref{e:24} above.

But the constraint for \eqref{e:24} can be removed via a modification to
the underlying platform.  The idea is to allow potentially any address
to stand for `the $n$th location on the stack'.  That requires an extra
hardware or software address translation unit between the runtime
software process and memory.  On being passed a new address $a$ intended
as the $n$th stack position, it remaps it to the next free address $b$
in a previously designated contiguous linear region of memory $R$, say
the region between 1GB and 1.25GB, and memoizes the choice so the next
time the address $a$ is passed to it, it is mapped to $b$ in $R$ again.
That gives each process access to 250MB, though the addresses $a$ it
generates range randomly across the full 32-bit range (0 to 4GB).  The
compiler inserts instructions in the object code that remove defunct
mappings, keeping the memory needed down.

That unit exists and is part of every processor.  It is the `translation
lookaside buffer' (TLB), and its job is to remap memory address space a
page (8KB) at a time, but we need it to work with individual addresses,
not whole pages.  An easy solution is to simulate it in an 
underlying software layer.  Given that,
the compiler is free to vary $n$ in \eqref{e:24} and the constraint in
\eqref{e:25} relates three constants, $k$, $n$ and ${\bf\Delta}n$,
allowing for $2\times$ 32 bits of entropy for the two `instructions
that write' in \eqref{e:24}, satisfying \eqref{hmax} and allowing
Theorem~\ref{t:2} to conclude the trace has been randomised (`trace
entropy is maximised').

But there are still other constraints on compiler-induced variation due to
computational semantics.
The inputs and outputs of a copy instruction are the same,
so the variations in them are the same and not independent.  Also, the
compiler must set the same variations from nominal values at the end
of a loop as at the beginning, because it cannot tell in general how
many times a loop will be traversed at runtime and must prepare the code
for another traversal after one time through.

We will not go through all the code constructs, just loops, to show how
and when the principle \eqref{hmax} must sometimes fail.

\subsection{Loops}

\noindent
Let the statement compiler $C[-]$ produce code ${\it mc}$ from
statement $s$ of the source language, changing the combined database
$D=({\bf\Delta},{\bf\eth})$ of offsets to $D^s$ in the process. The
offsets ${\bf\eth}$ are the intended variations from nominal values at runtime
for the data content of each register and memory location and the
offsets ${\bf\Delta}$ are the extra displacements for addressing 
described in Section~\ref{ss:ADC}.

For legibility, pairs $(D,x)$ will be written $D:x$ here. Then the
statement above of what the compiler does with statements $s$
is formalised as:
\begin{align}
D^s: {\it mc} &= C[D: s]
\label{e:26}
\end{align}
The notation emphasises the compiler is side-effecting on $D$.

Compiling {{\bf while}\,$e$\,$s$} means emitting code
{\it mc} constructed from ${\it mc}_e$ for $e$ and
${\it mc}_s$ for $s$, with this shape:
\begin{equation}
\begin{minipage}{0.5\columnwidth}
\begin{minipage}{1.5\textwidth}
\begin{code}
{\rm{start}}: {\it{mc}}\(\sb{e}\)       #{\rm compute \(e\) in \(r\)}
{\bf{beqz}} \(r\) {\rm{end}}      #{\rm goto to end if \(r\) zero}
{\it{mc}}\(\sb{s}\)       \,\,\,    #{\rm compute \(s\)}
{\bf{b}} {\rm{start}}     \,\,    #{\rm goto start}
{\rm{end}}:
\end{code}
\end{minipage}
\end{minipage}
\end{equation}
That does not work as-is, because the code does not at the end of the loop
reestablish the deltas that prevailed at loop start, so a second time through at
runtime, much goes wrong.
Extra code is needed after ${\it mc}_s$, so-called `trailer' instructions.

A trailer instruction adjusts the content of register $r$ back to the
delta ${{\bf{\eth}}} r$ off nominal at the beginning of the loop,
starting from a delta ${{\bf{\eth}}\sp{s}}r$ at the end of the loop.
It is as follows:
\begin{equation}
\begin{minipage}{0.5\columnwidth}
\begin{minipage}{1.5\textwidth}
\begin{code}
addi\,\(r\)\,\(r\)\,\(k\) \,\,     #{\rm where \(k={{\bf{\eth}}}\,r-{{\bf{\eth}}\sp{s}}\,r\)}
\end{code}
\end{minipage}
\end{minipage}
\label{e:28}
\end{equation}
Just the one instruction is required per register.

Trailer instructions that restore the $n$th stack location offset must also
restore the address displacement constant used in load and store.  That
is a more complex code sequence:
\begin{equation}
\hspace{0.15\columnwidth}
\begin{minipage}{0.6\columnwidth}
\begin{minipage}{1.33\textwidth}
\begin{code}
addi{\bf\,t0\,sp}\,\(j\)    \,#{\rm \(j=n-{\bf\eth} {\bf{sp}}-{\bf\Delta}\sp{s}n\)}
lw{\bf\,t0}\,\({\bf\Delta}\sp{s}n\)({\bf{t0}})  #{\rm load \(n\)th stack location}\vspace{0.75ex}
addi{\bf\,t0\,t0}\,\(k\)    \,#{\rm modify by \(k={{\bf{\eth}}n-{{\bf{\eth}}\sp{s}}}n\)}\vspace{0.75ex}
addi{\bf\,t1\,sp}\,\(l\,\)    #{\rm \(l=n-{\bf\eth} {\bf{sp}}-{\bf\Delta}n\)}
sw\,\({\bf\Delta}n\)({\bf{t1}}){\bf\,t0} \,\, #{\rm store \(n\)th stack location}
\end{code}
\end{minipage}
\end{minipage}
\label{e:29}
\end{equation}
(the $\bf t0$, $\bf t1$ registers are `temporary' workspace).
The first two instructions are the read stack code of
\eqref{e:24}, and the last two are the corresponding
write code. In-between, the instruction \eqref{e:28} changes the content
delta. The sequence reads with one address displacement, and
writes with another, changing the location of the (changed) content.

The trailer sequences (\ref{e:28}-\ref{e:29}) introduce no entropy at
all into the runtime trace and therefore fail the principle
\eqref{hmax} (effectively, per instruction that writes, 32 bits of
variation is needed from the compiler to mask programmer information).
The constants in the instructions are determined by choices of deltas by
the compiler for earlier instructions and it is impossible to execute
the trailer instructions without traversing the loop body, which will
execute those earlier instructions, so these trailer instructions are
always `old news' when they run. The content of the registers or memory
locations they affect during execution is correlated with the content
of registers and memory locations at the start of the loop.

Other places where the compiler must put trailer instructions like
(\ref{e:28}-\ref{e:29}) are where conditional branches join again, the
labelled targets of {\bf goto}s, and at {\bf return} from functions.
Calls of interior functions also require `trailers', but before the
call, because the delta offsets in force at the point where the function
was defined must be reestablished (that is a `come from' semantics; c.f.
{\bf goto}).  If there are any global variables, then the same holds 
of ordinary (i.e., not `interior') function calls, with respect to 
the global variables they access.

At each of those points the principle \eqref{hmax} fails, because every
instruction that writes (on a 32-bit platform) does not introduce 32
bits of entropy from the compiler.

\subsection{The Argument for \eqref{obfusc}}
\label{ss:proof}

\noindent
Given the analysis above, the argument for \eqref{obfusc}
requires a program to be unrolled far enough before chaotic compilation so none
of the points where \eqref{hmax} fails will be encountered in practice
by an observer.  Then by Theorem~\ref{t:1} the entropy in the observed
part of the trace is sufficient to mask completely any input from the
programmer, and an observer cannot say with any significant probability
of being right (as the word length $n$ tends to infinity) what any bit
of data picked out of the trace means.
A fuller sketch of the argument is as follows:

\begin{proof}[Sketch for \eqref{obfusc}]
Suppose for contradiction that the observer has a polynomial time method
$f(n,T)$ of working out what the data beneath the encryption is at some
$m$ points of interest in the trace $T$ of program $P$, where $n$ is the
platform word size.  For brevity assume $P$ includes its intended
inputs, compiled-in.  The observer sees runtime trace $T$.

WLOG take $m{=}1$ and suppose the observer is interested in only one
particular bit in that particular register or location in memory at
that point, say the least significant bit $b$.  It is last written to in
the $M$th step of the trace before the point of interest.  Then the
observer's prediction $f(n,T)$ is 0 or 1.  Suppose also the program has
been written in the machine code subset consisting of addition
instructions with semantics $x \,{\leftarrow}\,y {+} k$ (${\bf
addi}\,x\,y\,k$) for a constant $k$ embedded in the instruction, and
branch instructions with semantics $\IF x {<}y {+}k \, \keyword{goto}\,
L$ (${\bf blt}\,x\,y\,k\,L$) for constant $k$ and target address $L$
embedded in the instruction.  (Those are enough for computable
functions, i.e., `programs'; c.f., the programming language Fractran
\cite{conway87fractran}.)

A challenger readies a sequence of maximal entropy
compilations $C[P_n]$ of $P$ with the $n$th being for a $n$-bit
platform as target, and $P$ having been partially or completely unrolled
as $P_n$ with no loops or branches in the first $2^n$ (i.e.,
super-polynomially many) machine code instructions.  If the program
predictably ends before then, it is to be unrolled completely.  These
are compilations of the same program $P$ all with the same end-to-end
semantics that could be produced entirely automatically  
(the constants in $P$ can be expressed exactly in the
$n$ bit words available on the platform, for all $n$ considered). The observer
is invited to apply their method $f$ and predict the chosen bit $b$ in the
traces $T_n$.  By hypothesis, $f(n,T_n)$ is correct with probability at
least $B$ with $B>1/2$ as $n\to\infty$.

Let $N$ be such that for $n\ge N$ 
the word of interest for the observer in
the trace is produced by the unrolled part, that is, $2^N\ge M$.
Consider the program $Q$ that is $P_N$ with
the instruction that produces the word of interest changed from 
$x \,{\leftarrow}\,y {+} k$ to $x \,{\leftarrow}\,y {+} k'$ where
$k'=k{+}1$, so the word is written in $Q$ to 1 more than in $P_N$.
But also all instructions in $P_N$ that read that register or memory
location $x$ before it is written over again are changed in $Q$ to compensate.
That is, an instruction $z \,{\leftarrow}\,x {+} k_1$ in $P_N$ is changed to
$z\,{\leftarrow}\,x {+} k_1'$ in $Q$,  where $k_1'=k_1{-}1$; an instruction
$\IF x {<}z {+}k_2 \, \keyword{goto}\, L$ in $P_N$ is changed to 
$\IF x {<}z {+}k_2' \, \keyword{goto}\, L$ in $Q$, where $k_2'=k_2{-}1$.
(There are only a finite number of instructions to change, as $P_N$ is
finite, but they may appear infinitely many times in the trace.)

Then the trace of $Q$ is the same as the trace of $P_N$ (and $P_n$, for
$n\ge N$ except for that
one word and its least significant bit. Moreover, the traces from the
code $C[Q]$ are each of them traces that may have been produced from
code $C[P_N]$ (and/or any $P_n$ for $n\ge N$) since the change from 
$P_N$ to $Q$ is one that the compiler may make, i.e. $Q=C[P_N]$ is
possible.

Then the method $f(n,-)$ applied to a trace of $C[Q]$ on a platform
with an $n$ bit word produces the same result as when applied to a 
trace of $C[P_N]$ (and $C[P_n]$ for $n\ge N$), and so must produce
the right result for $P$ with probability at least $B>1/2$ for
$n$ large enough. But that is the wrong result for $Q$. So the method
$f(n,-)$ produces the right result for $Q$ with probability $1-B < 1/2$
for large enough $n$, contradicting the hypothesis, so the 
hypothetical method $f$ does not exist. \hfill \QED
\end{proof}

\smallskip

The result for $m{\ge}1$ follows because if the observer had a method
$f$ that made predictions at $m{>}1$ points in the trace,
then the method $g$ that throws away the $m{-}1$ predictions of $f$'s at
unwanted points would work as a method for $m{=}1$, and the proof rules
that out. Similarly for more than one bit.

The argument is not deep, as the empty program case shows, but it is
hard to codify and should increase understanding of
this area.  The chaotic compiler constructions are reminiscent of Yao's {\em
garbled circuits} \cite{yao86}, at $n$ bits and with recursion.

\section{Implementation}
\label{s:Imp}

\def\ME[#1]{#1}
\def\la{\(\leftarrow\)}
\begin{table}[!tbp]
\caption{Trace for Ackermann(3,1)}
\label{tab:3}
\flushleft
\begin{tabular}{@{}l@{}r@{}}
\begin{minipage}{0.6\columnwidth}
\begin{alltt}\scriptsize
{\rm{PC}}  {\rm{instruction}}                     {\rm{trace updates}}
\dots
35  addi t0 a0      \ME[-86921031]       t0 \la \ME[-86921028]
36  addi t1 zer     \ME[-327157853]      t1 \la \ME[-327157853]
37  beq  t0 t1  2   \ME[240236822]                  
38  addi t0 zer     \ME[-1242455113]     t0 \la \ME[-1242455113]
39  b 1                                             
41  addi t1 zer     \ME[-1902505258]     t1 \la \ME[-1902505258]
42  xor  t0 t0  t1  \ME[-1734761313] \ME[1242455113] \ME[1902505258]
                                    t0 = \ME[-17347613130]
43  beqz t0 9       \ME[-1734761313]                 
53  addi sp sp      \ME[800875856]       sp \la \ME[1687471183] 
54  addi t0 a1      \ME[-915514235]      t0 \la \ME[-915514234] 
55  addi t1 zer     \ME[-1175411995]     t1 \la \ME[-1175411995]
56  beq  t0 t1  2   \ME[259897760]                   
57  addi t0 zer     \ME[11161509]        t0 \la \ME[11161509]   
\dots
143 addi v0 t0      \ME[42611675]       \fbox{v0 \la \ME[13]} #{\rm result}
\dots
147 jr  ra
STOP
\end{alltt}
\end{minipage}
\\
\\
\begin{minipage}[t]{1.0\columnwidth}
\scriptsize%
\begin{tabular}{@{}|@{\,}l@{\,}l@{\,}l@{\,}|@{\,}ll@{\,}|@{}}
\hline
\multicolumn{5}{@{}|c|@{}}{\vbox to 3ex{}Legend}\\
\hline
\em op.& {\em fields} & {\em semantics} & {\em register} & {\em use}\\
\hline
\vbox to 3ex{}addi &$\rm r_0$ $\rm r_1$ $k$ 
        &$r_0 \leftarrow r_1+k$
        & {a0,a1,\dots} & function argument
        \\
b   &$i$
        &${\it pc}\leftarrow {\it pc}+i$
        & {pc} & program counter
        \\
beq& $\rm r_1$ $\rm r_2$ $i$
        &${\rm if} r_1 {=} r_2 {\rm then} {\it pc}{\leftarrow}{\it pc}{+}i$
        & {ra} & return address
        \\
jr   &$\rm r$
       &${\it pc} \leftarrow r$
        & {sp} & stack pointer
       \\
xor &$\rm r_0$\kern1pt$\rm r_1$\kern1pt$\rm r_2$
        &$r_0\leftarrow (\kern-1ptr_1{+}k_1)\mathop{\widehat{ }}(\kern-1ptr_2{+}k_2){-}k_0$
        &{t0,t1,\dots} & temporaries
        \\
         & \kern1pt$k_1$\kern1pt$k_2$\kern1pt$k_0$&
             &{v0,v1,\dots} & return value
                \\
\hline
\hline
\multicolumn{5}{@{}|l|@{}}{{$i$} program count increment,
                      {$k$} instruction constant,
                      {$r$} content of r}\\
\hline
\end{tabular}
\end{minipage}
\end{tabular}
\end{table}
 
\noindent
A prototype C compiler \url{http://sf.net/p/obfusc} 
covers {\sc ansi} C and GNU C extensions, including
statements-as-expressions and expressions-as-statements, gotos, arrays,
pointers, structs, unions, floating point, double integer and floating
point data.  It is missing {\bf longjmp} and efficient strings
({\bf char} and {\bf short} are the same size as {\bf int}).
It is intended for use in encrypted computing
\cite{fletcher2012,BB13a,oic,heroic,cryptoblaze18,BB18d},
an emerging processor technology in which inputs, outputs, and all
intermediate values in registers and memory are in encrypted form.
Because good encryption is one-to-many, and addresses are data like any
other, many physically different bit sequences (the `ciphertext') 
represent the address intended by the programmer (the `plaintext'), and
platforms exhibit hardware aliasing at every memory access, providing a
testbed for theory and practice.

A trace\footnote{Initial and final content offset deltas
are set to zero here, for readability.} of the Ackermann
function\footnote{Ackermann
C code: {\bf int} A({\bf int} m,{\bf int} n) \{ {\bf if} (m == 0) {\bf
return} n+1; {\bf if} (n == 0) {\bf return} A(m-1, 1); {\bf return}
A(m-1, A(m, n-1)); \}.} \cite{Sundblad71} compiled by the compiler
is shown in Table~\ref{tab:3}.  The Ackermann function is the most
computationally complex function possible, increasing in the degree of
complexity required to calculate it (polynomial, exponential,
super-exponential, etc.) with each increment of the first argument.  It
is not practically possible to `cheat' in the calculation, and the
compilation exercises the basic code constructs (function call,
conditional, arithmetic, etc.), and the calculation is very sensitive,
so it is a stiff test.  The trace illustrates how the compiler's
variation of the delta offsets for register content through the code
results in randomly generated constants embedded in the instructions and
randomly offset runtime data -- until the result is returned with offset 0.

\begin{table}[!tp]
\caption{Trace for sieve showing extra bits (right) ignored
in arithmetic but significant in memory addressing.
Stack read lines gray, address base red, displacement violet.
}
\def\comm#1{}
\label{tab:4}
\flushleft
\begin{tabular}{@{}l@{}r@{}}
\begin{minipage}{0.98\columnwidth}
\begin{alltt}\scriptsize
{\rm{PC}}     {\rm{instruction}}                         {\rm{trace updates}} | {\rm{extra bits}}
\dots
22340 addi t1  sp  \ME[-418452205]    t1 \la \ME[-877254954|1532548040]
22360 bne  t0  t1  84
{\color{gray}22384 addi t1  sp  \ME[-407791003]    t1 \la \ME[{{\color{red}-866593752|1532548040}}]}\comm{\rm read local array}
{\color{gray}22404 lw   t0  \ME[{{\color{violet}866593746}}](t1)     t0 \la \ME[-866593745|1800719299]}\comm{a[7]{\rm at}{\it sp}{\rm+40}}
22424 addi t0  t0  \ME[-1668656853]   t0 \la \ME[1759716698|1081155516]
22444 b    540
22988 addi t1  zer \ME[1759716697]    t1 \la \ME[1759716697|1325372150]
23008 bne  t0  t1  44
\dots
{\color{gray}23128 addi t0  sp  \ME[-1763599776]   t0 \la \ME[{{\color{red}2072564771|-1935092797}}]}\comm{\rm read local variable}
{\color{gray}23148 lw   t0  \ME[{{\color{violet}-2072564772}}](t0)   t0 \la \ME[2072564779|-1773201679]}\comm{\rm at{\it sp}{\rm+45}}
23168 addi t0  t0  \ME[1723411350]    t0 \la \ME[-498991167|-981581771]
23188 addi t0  t0  \ME[-1862832992]   t0 \la \ME[1933143137|-1629507929]
23208 addi v0  t0  \ME[-1933143130]  \fbox{v0 \la \ME[7]}        \:|1680883739 \comm{ return}
\dots
23272 jr   ra
STOP
\end{alltt}
\noindent
{\rm{See Table\,\ref{tab:3} for Legend.}}
\end{minipage}
&
\end{tabular}
\end{table}

Running a Sieve of Eratosthenes program\footnote{Sieve C code:
{\bf int}
S({\bf int} n ) \{ {\bf int} a[N]=\{[0\dots N-1]=1,\};
    {\bf if} (n$>$N$||$n$<$3) {\bf return} 0;
    {\bf for} ({\bf int} i=2; i$<$n; ++i) \{
        {\bf if} (!a[i]) {\bf continue};
        {\bf for} ({\bf int} j= 2*i; j$<$n; ++j) a[j]=0;
        \}; 
    {\bf for} ({\bf int} i=n-1; i$>$2; {-}{-}i)
        if (a[i]) {\bf return} i;
    {\bf return} 0;
\}
.
}
for primes is a delicate test of memory-oriented programming.
The final part of the trace  is shown in Table~\ref{tab:4} with
two stack reads in gray and the address base in red, with the address
displacement (in the instruction) in violet. The trace shows `extra bits'
that form part of the data (and addresses) but which are not used
in the arithmetic, in order to generate hardware aliasing. The extra
is formed deterministically by hashing the extra bits and visible
data of the operands to each arithmetic operation.

\section{Conclusion}

\noindent
This paper has described the compilation of imperative source for a
platform that has hardware aliasing with hidden determinism.  The
technique depends on the compiler controlling the address displacement
and address base for load and store instructions so that they are always
the same for repeat accesses to the same memory location.  That 
uses saved copies or repeats earlier calculations exactly.  The
technique is extended to also generate as well as compensate for random
displacements in the addressing, and extended again to generate and
compensate for random displacements in the data content of all registers
and memory (called chaotic compilation).  A sketch proof is included to
show this technique results in object code where it is difficult to
discover the programmer's intention at any points of the runtime trace,
while maintaining the semantics.  I.e., the compiled code is
technically `obfuscated'.

\begin{small}
\bibliography{issre2019-nonanon}
\end{small}

\end{document}